\pdfoutput=1

\documentclass[11pt]{article}

\usepackage[]{style/ACL2023}

\usepackage{times}
\usepackage{latexsym}

\usepackage[T1]{fontenc}

\usepackage[utf8]{inputenc}

\usepackage{microtype}
\usepackage{booktabs}
\usepackage{multirow}
\usepackage{amsmath}
\usepackage{amssymb}
\usepackage{graphicx}

\usepackage{pifont}

\usepackage[breaklinks]{hyperref}

\usepackage{inconsolata}

\usepackage{paralist}
\usepackage{url}

\title{Embedding Mental Health Discourse for Community Recommendation}

\author{Hy Dang$^*$, Bang Nguyen$^*$, Noah Ziems, Meng Jiang \\
  University of Notre Dame \\
  \texttt{\{hdang, bnguyen5, nziems2, mjiang2\}@nd.edu}
}

\begin{document}
\maketitle
\def\thefootnote{*}\footnotetext{These authors contributed equally to this work}
\begin{abstract}

Our paper investigates the use of discourse embedding techniques to develop a community recommendation system that focuses on mental health support groups on social media. Social media platforms provide a means for users to anonymously connect with communities that cater to their specific interests. However, with the vast number of online communities available, users may face difficulties in identifying relevant groups to address their mental health concerns. To address this challenge, we explore the integration of discourse information from various subreddit communities using embedding techniques to develop an effective recommendation system. Our approach involves the use of content-based and collaborative filtering techniques to enhance the performance of the recommendation system. Our findings indicate that the proposed approach outperforms the use of each technique separately and provides interpretability in the recommendation process.

\end{abstract}

\section{Introduction}
\label{sec:intro}

The rise of social media as a platform has allowed people all over the world to connect and communicate with one another.
Further, these communities that exist online are able to keep their members anonymous from one another, allowing new communities to form which would have a hard time existing without anonymity.

Specifically, this new and robust anonymity has allowed an explosion of online communities with a focus on giving each other advice on health issues.
While being involved in seeking peer support in a community with people that have experienced similar issues can provide a significant positive impact on someone's ability to navigate their personal problems \cite{richard2022scoping}, finding communities with relevant discourse is not trivial.
Often, the platforms which host these communities have a very large quantity of them.
There are over 100,000 different communities on Reddit alone.
Further, some communities are not easily found due to their inherently anonymous nature, so the only way a user can decide if they fit within the community is by spending time reading through the discourse happening within the community.

For these reasons, new users seeking others who have experienced similar situations may have a very hard time finding communities that would help them the most, even if they are familiar with the platform which hosts the communities.

Recently, embedding long sequences of text has received lots of interest both from the research community and from practitioners.
A number of studies have shown embeddings can be useful for measuring the similarity both between document pairs and between question-document pairs \cite{karpukhin2020dense, xiong2020approximate, qu2021rocketqa}, allowing for retrieval of the most similar documents given a new question or document.
However, little work has been done investigating how the discourse within a community, which represents the meaning of that community, can be represented in a single embedding. The discourse of a community in this context can be all users' posts in that specific community or represented community's description. 
This poses a unique challenge as discourse within these communities is often in the form of threads that, unlike documents, are not naturally represented as a single block of text.

The goal of this work is to develop a system to recommend support groups to social media users who seek help regarding mental health issues using embeddings to represent the communities and their discourse.
Specifically, we aim to leverage the text of a given user's posts along with the description and posts in each subreddit community to help recommend support groups that the user could consider joining.

Our main research questions are as follows:
\begin{enumerate}
\item In representing online communities through discourse embeddings, what type of information can be used?
\item To what degree do these representations improve the accuracy of predicting users' behaviors regarding their involvement in sharing experiences within groups or communities?
\item Do different discourse embedding methods change the prediction capacity of our community recommendation model?
\end{enumerate}

In exploring these research questions, we propose a hybrid recommendation approach that leverages both content-based and collaborative filtering to construct our community recommendation model. As shown in Fig. \ref{fig:workflow}, the content-based filtering component investigates different methods of embedding discourse within a community to recommend similar communities to users. It is then combined with a matrix factorization model that learns user engagement behavior in a community to improve recommendation decisions. Utilizing users' past interactions as well as text-based information about the communities, we show that our model achieves promising accuracy while offering interpretability.

\section{Related Work}
\label{sec:related}

There are a number of studies related to our work.

\citet{son2022discourse} and \citet{balusu2022pretrained} constructed discourse embeddings to find relations between short text segments.
While the two studies were similar in concept, they focused on short text segments where this work instead focused on constructing discourse embeddings for entire social media communities.

\citet{garriga2022machine} showed NLP techniques could be used with electronic health records to predict mental health crises 4 weeks in advance.
While online communities were no replacement for professional medical help, this suggested many who had looming mental health problems seek help before a crisis.

\citet{low2020natural} experimented on the same dataset we used with Natural Language Processing techniques such as TF-IDF and sentiment analysis to understand the effects of COVID-19 on mental health.
Although working on the same dataset, our work studies a different task: to recommend mental health-related support community to Reddit users. 

\citet{musto2016learning} adopted a similar approach to ours in content-based filtering for recommendation.
Specifically, they mapped a Wikipedia page to each item and generate its corresponding vector representation using three feature-extraction methods - Latent Semantic Indexing, Random Indexing, and Word2Vec.
We extended this method by exploring more recent representations of text such as BERT \cite{bert} and OpenAI embeddings.

\citet{halder2017health} recommended threads in health forums based on the topics of interest of the users.
Specifically, self-reported medical conditions and symptoms of treatments were used as additional information to help improve thread recommendations~\citep{wang2020calendar,jiang2012social}.
While our work is also situated in the health domain, we are interested in recommending a broader support group to users rather than a specific thread. 

\citet{ghazarian2022wrong} used sentiment and other features to automatically evaluate dialog, showing NLP techniques could be used to evaluate quality of discourse.
In doing so, they leveraged weak supervision to train a model on a large dataset without needing quality annotations.

\section{Problem Definition}
\label{sec:problem}

Suppose we have a Reddit's "\textit{who-posts-to-what}" graph, which is denoted by  $G = (U, V, E)$ where $U$ is the set of users, $V$ is the set of subreddit communities, and $E$, a subset of $U\times V$, is the set of edges. 
The number of user nodes is $m = |U|$ and the number of subreddit communities is $n = |V|$. So, $U = \{(u_1, P_1), (u_2, P_2) , ..., (u_m, P_m)\}$ where $P_i$ is the set of posts by user $u_i$ and $V = \{(v_1, P^{\prime}_1), ..., (v_n, P^{\prime}_n)\}$ where $P^{\prime}_j$ is the set of all posts in subreddit $v_j$.
If a user $u_i$ posts to subreddit $v_j$, there is an edge that goes from $u_i$ to $v_j$, which is denoted by $e_{ij} = e(u_i, v_j)$. 
The problem is that given $G$, predict if $e_{ij} = e(u_i, v_j)$ exists.
In other words, will user $u_i$ post to subreddit $v_j$?

\section{Methodology}
\label{sec:method}
\begin{figure*}[ht]
    \centering
    \includegraphics[width=0.99\textwidth]{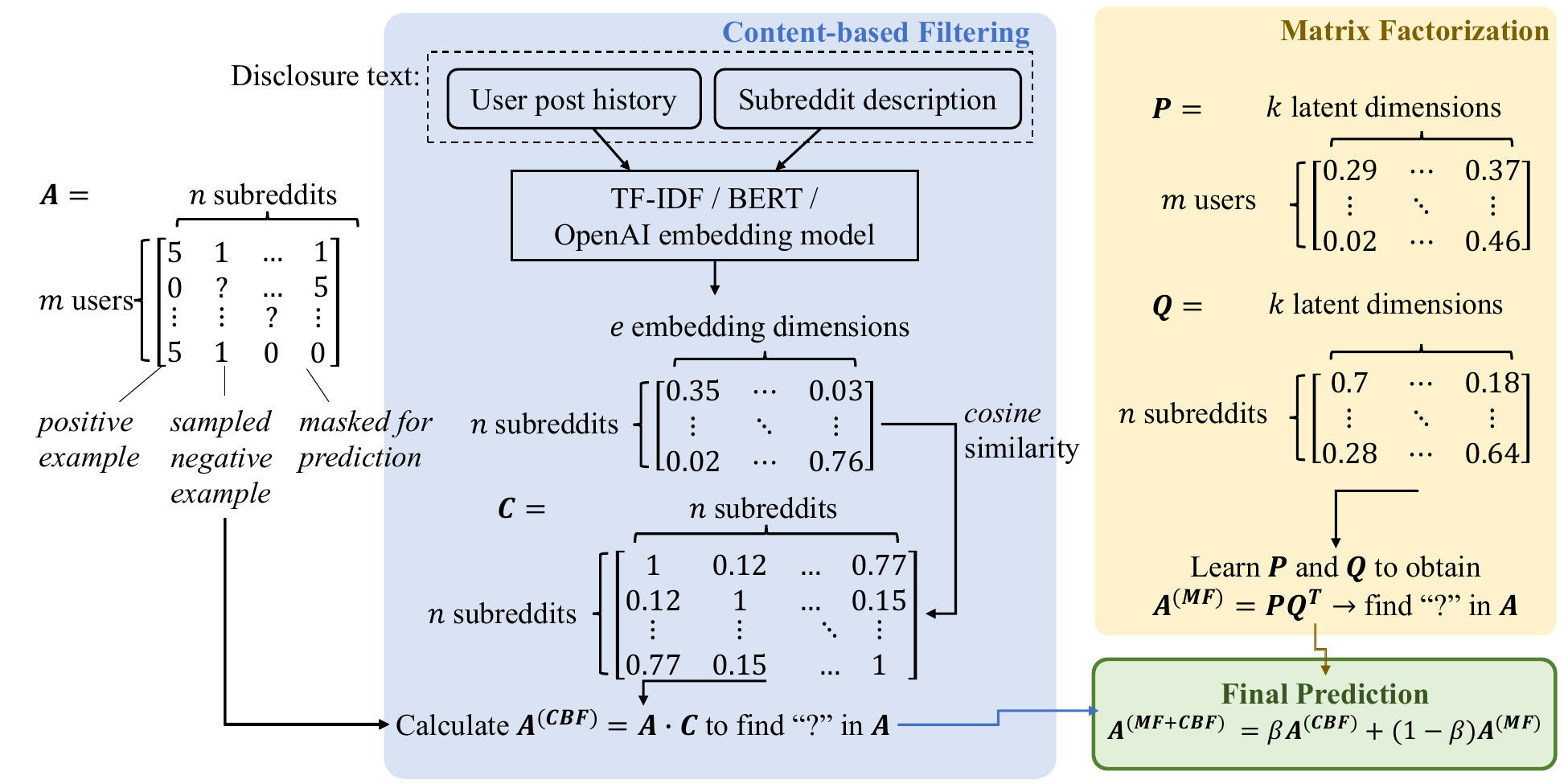}
    \caption{Our recommendation pipeline, which linearly combines the prediction of a content-based filtering (CBF) and a matrix factorization (MF) model. In the CBF model, recommendations of new subreddits are made through the average of a user's past interaction, weighted by how similar the past subreddits are to the new ones. In the MF model, users and subreddits are represented in a joint latent space of $k$ dimensions. Recommendations of new subreddits are made based on the distance between users and subreddits in this latent space.}
    \label{fig:workflow}
\end{figure*}
Figure \ref{fig:workflow} illustrates our recommendation pipeline, which adopts a hybrid approach by incorporating both content-based filtering (CBF) and collaborative filtering, specifically matrix factorization (MF) strategies. The CBF model recommends new subreddits based on the average of a user's previous interactions, weighted by how similar the previous subreddits are to the new ones. Meanwhile, users and subreddits are represented in a $k$-dimensional joint latent space in the MF model. The distance between users and subreddits in this latent space is used to provide recommendations for new subreddits. The predictions from these two components are linearly combined to obtain the final recommendation of subreddits to users.

The collaborative filtering component of our solution leverages nonnegative matrix factorization to represent our users and subreddits in lower-dimensional latent space. In this sense, we redefine the adjacency matrix $\mathbf{A}$ in our problem definition so that it works with nonnegative factorization. More specifically, users' past interactions with items are represented by the adjacency matrix $\mathbf{A} \in {\{5, 1, 0\}}^{m \times n}$. $A_{ij} = 5$ if the user $u_i$ has posted to subreddit $j$, $A_{ij} = 1$ if the user $u_i$ has NOT posted to the subreddit $v_j$, and $A_{ij} = 0$ is the missing connection that needs predicting. Given this adjacency matrix $\mathbf{A}$, the task is to predict the missing elements $A_{ij} = 0$. In the following sections, we elaborate on each component of our recommendation model and then discuss how they are combined to obtain our final solution.

\subsection{Content-based Filtering}

 In recommending items to users based on their past interactions and preferences, content-based filtering methods represent each item with a feature vector, which can then be utilized to measure the similarity between items \cite{content-filtering}. If an item is similar to another item with which a user interacted in the past, it will be recommended to that same user. Thus, in addition to the adjacency matrix $\mathbf{A}$, we utilize another matrix $\mathbf{C}$ of size $n\times n$, where $\mathbf{C}_{ab}$ is the similarity between the embeddings for two subreddits with embedding vectors $\mathbf{a}$ and $\mathbf{b}$.
In this paper, we use cosine similarity as the similarity measure:
\[
\mathbf{C}_{ab} =  \dfrac {\mathbf{a} \cdot \mathbf{b}} {\left\| \mathbf{a}\right\| \left\| \mathbf{b}\right\| },
\]

To predict the value of the missing element where $A_{ij} = 0$ (whether user $u_i$ will post to subreddit $v_j$), we compute the average of user $u_i$'s past interactions (which subreddits user $u_i$ posted and did not post to), weighted by the similarity of these subreddits to subreddit $v_j$.
Mathematically,
\[
   {A}^{\prime}_{ij} =  \frac {\sum_{k=1}^{n} A_{ik} C_{kj}}{\sum_{k=1}^{n} C_{kj}}.
\]
We can generalize the above formula to obtain the new predicted adjacency matrix using matrix-level operations:
\[
   \mathbf{A}^{\text{(CBF)}} =  (\mathbf{A}\mathbf{C}) \odot \mathbf{D},
\]
where
\begin{compactitem}
    \item $\mathbf{D} = 1. / (\mathbf{I} \cdot \mathbf{C})$ (element-wise),
    \item $\mathbf{I}$ is an indicator matrix such that ${I}_{ij} = 1$ if ${A}_{ij} \neq 0$, otherwise ${I}_{ij} = 0$,
    \item and $\odot$ is the Hadamard product.
\end{compactitem}

\subsubsection{Representing Subreddit Discourse with Description and Posts}
It is helpful to consider the specific domain of the application to represent each item as an embedding. In the context of our subreddit recommendation problem, we take advantage of two types of text-based information about a subreddit to construct the similarity matrix:  (1) the posts within the subreddit itself and (2) the general description about the reddit provided by the subreddit moderators.

We then use a feature extraction method to obtain two embeddings of a subreddit, one based on its description and the other based on its posts. As a subreddit contains many posts, each of which has a different embedding given the same feature-extraction method, we take the average of the embeddings across all posts within a subreddit to obtain one embedding for the subreddit.

\subsubsection{Feature Extraction}
In this paper, we consider three feature-extraction methods: {Term Frequency-Inverse Document Frequency}  (TF-IDF), {Bidirectional Encoder Representations from Transformers} (BERT) \cite{bert}, and {OpenAI}.\footnote[1]{OpenAI API Embeddings: \url{https://platform.openai.com/docs/guides/embeddings}}

\textbf{TF-IDF}: The TF-IDF algorithm represents a document as a vector, each element of which corresponds to the TF-IDF score of a word in that document.
The TF-IDF score for each word in the document is dictated by (1) the frequency of the word in the document \cite{idf}, and (2) the rarity of the word in the entire text corpus \cite{tf}.
That is, a term is important to a document if it occurs frequently in the document but rarely in the corpus.
We use the implementation from scikit-learn \cite{scikit-learn} to obtain the TF-IDF representations of our subreddits.

\textbf{BERT}: We employ BERT to generate sentence embeddings as another feature extraction technique \cite{bert}.
BERT takes a sentence as input and generates a fixed-length vector representation of the sentence.
This representation is meant to capture the syntactic and semantic meaning of the input sentence in a way that can be used for various natural language processing tasks, such as sentence classification or semantic similarity comparison.
In the context of our problem, we can treat each subreddit description or each post as a sentence and feed it to a pre-trained BERT model to generate the embeddings that represent the subreddit. Long posts are truncated to fit within the context limits of pre-trained models. We experiment with 4 different variations of BERT embeddings:
\begin{itemize}
    \item BERT base and large \cite{bert}
    \item Sentence-BERT, or SBERT \cite{reimers2019sentence}
    \item BERTweet \cite{bertweet} 
\end{itemize}
\textbf{OpenAI}: Similar to BERT embeddings, OpenAI embeddings take in a string of text and output an embedding that represents the semantic meaning of the text as a dense vector.
To do this, the input string is first converted into a sequence of tokens.
The tokens are then fed to a Large Language Model (LLM), which generates a single embedding vector of fixed size.
OpenAI's text-embedding-ada-002 can take strings of up to 8191 tokens and returns a vector with 1536 dimensions.

\subsection{Nonnegative Matrix Factorization for Collaborative Filtering}

Matrix factorization (MF) approaches map users and items (subreddits in this case) to a joint latent factor space of a lower dimension $k$ \cite{koren2009matrix}. The goal of this method is to recommend to a user the subreddits that are close to them in the latent space. More formally, MF involves the construction of user matrix $\mathbf{P}$ of dimension $m\times k$ and subreddit matrix $\mathbf{Q}$ of dimension $n\times k$. In this sense, the resulting term, ${\mathbf{p}_i}^{\top} {\mathbf{q}_j}$, captures user $u_i$'s interest in item $v_j$’s characteristics, thereby approximating user $u_i$'s rating of item $v_j$, or denoted by ${A}_{ij}$.

This modeling approach learns the values in $\mathbf{P}$ and $\mathbf{Q}$ through the optimization of the loss fuction

\[
   \min_{\mathbf{P},\mathbf{Q}} \sum_{A_{ij} \in \mathbf{A}} ( A_{ij} - \mathbf{p}_{i}^{\top} \mathbf{q}_{j} )^{2} + \lambda ( \left\| \mathbf{p}_i \right\| ^2 + \left\| \mathbf{q}_j \right\|^2).
\]

Matrix factorization offers the flexibility of accounting for various data and domain-specific biases that may have an effect on the interaction between user $u_i$ and subreddit $v_j$. In this paper, we consider three types of biases: global average $\mu$, user bias $b_{i}^{(p)}$, and subreddit bias $b_{j}^{(q)}$. The updated loss function is given by:

\begin{equation}
\begin{split}
   \min_{\mathbf{P},\mathbf{Q}} \sum_{A_{ij} \in \mathbf{A}} ( A_{ij} - \mu - b_{i}^{(p)} - b_{j}^{(q)} - \mathbf{p}_{i}^{\top} \mathbf{q}_{j} )^{2} + \\  
   \lambda ( \left\| \mathbf{p}_i \right\| ^2 + \left\| \mathbf{q}_j \right\|^2 + b_{i}^{(p)^2} + b_{j}^{(q)^2}).
\end{split}
\end{equation}

After optimization, each element in the new predicted adjacency matrix $\mathbf{A^{\text{MF}}}$ is given by:
\[
   \mathbf{A}^{\text{(MF)}}_{ij} = \mathbf{p}_{i}^{\top} \mathbf{q}_{j} + \mu + b_i + b_j
\]

\subsection{Final Model: Hybrid Approach}
Our main model leverages insights from both content-based filtering and matrix factorization by taking a linear combination of their predicted adjacency matrix. Specifically, the new adjacency matrix is given by:
\[
   \mathbf{A}^{\text{(MF+CBF)}} = \beta \mathbf{A}^{\text{(CBF)}} + (1 - \beta) \mathbf{A}^{\text{(MF)}},
\]
where $\beta$ is a hyperparameter that controls how much the CBF model (vs MF model) contributes to the final prediction.

\section{Data and Experimental Setup}
\label{sec:experiment}
For the experimental setup, we use the data from \citet{low2020natural} working on Reddit platforms in mental health domains, particularly health anxiety.

\subsection{Data Description}
The dataset is collected from 28 mental health and non-mental health subreddits.

The dataset is suitable for studying how subreddits and social media platforms correlated with individuals' mental health and behavior.
The original data comprises 952,110 Reddit posts from 770,176 unique users across 28 subreddit communities, which include 15 mental health support groups, 2 broad mental health subreddits, and 11 non-mental health subreddits. We also manually collect descriptions of the 28 subreddits and use that information along with the posts to conduct the content similarity matrix.

\subsection{Data Preprocessing}
\begin{figure}[ht]
  \centering
  \includegraphics[width=80mm]{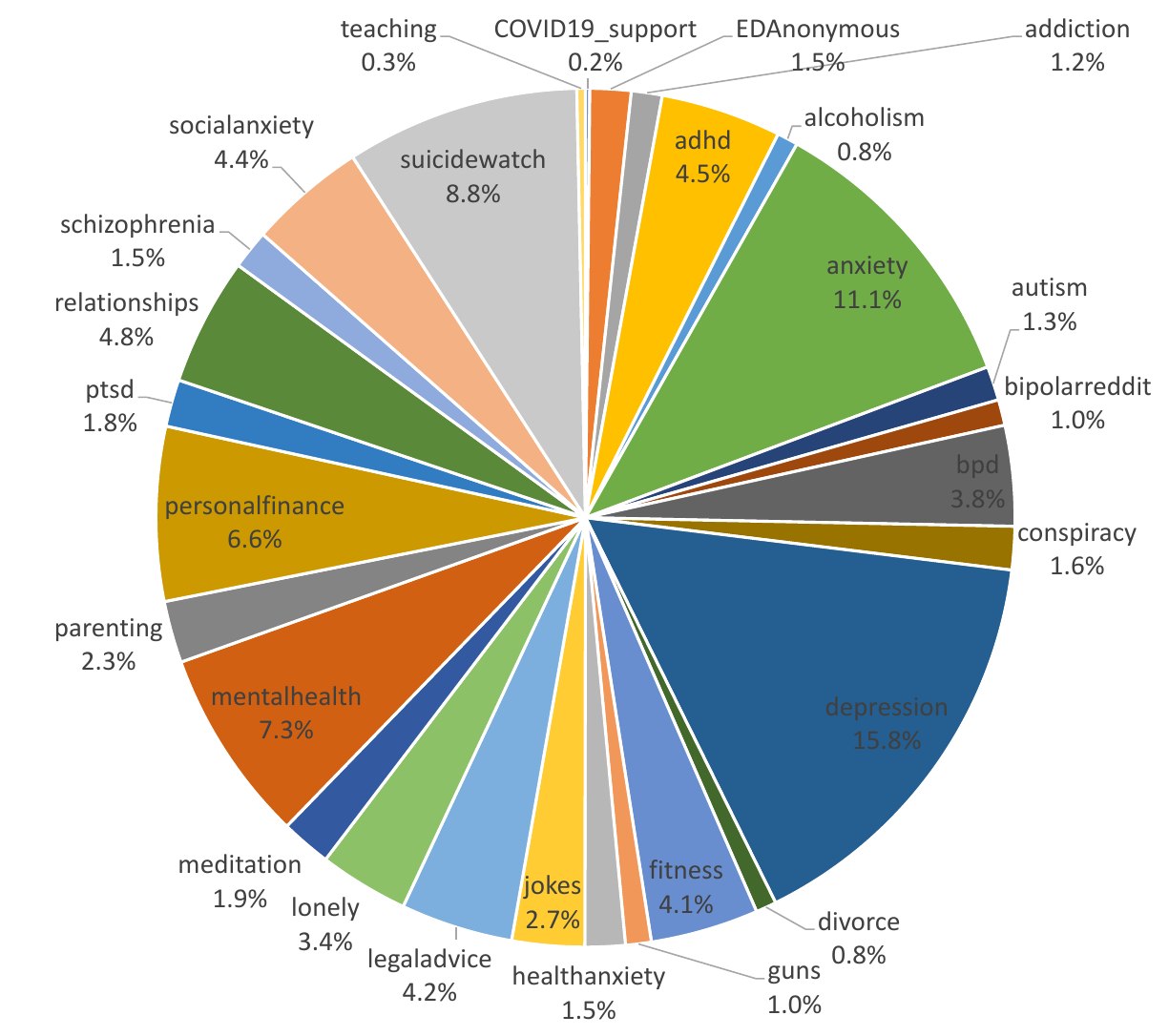}
  \caption{Distribution of post frequency across subreddits: r/depression, r/anxiety, and r/suicidewatch are the three most popular subreddits.}
  \label{frequentpost}
\end{figure}

Although the original dataset has a large number of unique users, the majority of them only contribute posts to one or two different communities. This presents a challenge when evaluating our specific task. As our objective is to examine users' behavior over time and provide recommendations for engaging in suitable subreddits, we have implemented a filter to exclude users who post to fewer than three subreddits. After filtering, the remaining users and posts are 16,801 and 69,004, respectively, while the number of subreddits remains to be 28. 
We also seek to understand the distribution of interactions between users and different subreddits. The detailed distribution of post frequency across subreddits is visualized in Figure~\ref{frequentpost}.

\subsection{Experimental Setup}
\subsubsection{Data Splits}
To construct our data splits, for each user in our dataset, we choose the most recent subreddit that the user first posted to as the test example.
For example, if the user post history is [\textit{subreddit1, subreddit2, subreddit3, subreddit1, subreddit2}], then \textit{subredddit3} will be used as the test example.
For each positive training example, we pair it with a negative example randomly sampled from the list of subreddits where the user has not posted to.

\subsubsection{Evaluation Metrics}
In assessing the performance of our recommendation method and the baseline, we use the following evaluation metrics: $Recall@K$ and \textit{Mean Reciprocal Rank (MRR)}.

\subsection{Results}

\begin{figure}[t]
  \centering
  \includegraphics[width=75mm]{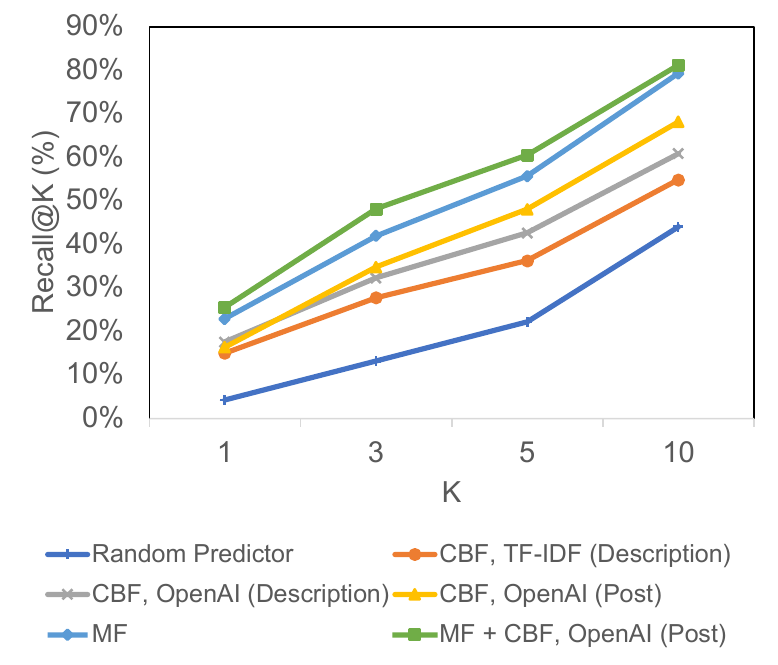}
  \caption{Model Performance on Recall@K}
  \label{hit_at_k}
  \vspace{-0.1in}
\end{figure}

Table \ref{result} presents the performance of our hybrid recommendation system as well as its individual components (MF or CBF). For CBF, we report its performance on different types of embeddings constructed using different information (posts or description) and different feature extraction methods (TF-IDF, BERT, or OpenAI). Figure~\ref{hit_at_k} visualizes the results of exemplary models in a diagram for better analysis using Recall@K.

According to Table \ref{result}, all variants of our recommendation method outperform the random predictor. Among all the variants, the hybrid solution using the content similarity matrix generated from OpenAI embeddings achieves the highest performance in MRR (0.4244) and average Recall@K. 

\begin{table*}[t]
\begin{center}
\begin{tabular}{|l|c|c|c|c|c|}
\hline \bf Approach & $\mathbf{MRR}$& $\mathbf{Recall@1}$ & $\mathbf{Recall@3}$ & $\mathbf{Recall@5}$& $\mathbf{Recall@10}$ \\ \hline 
\hline
Random Predictor & $0.1631$ &$0.0429$ &$0.1318$& $0.2221$ & $0.4409$ \\
\hline
\hline
Matrix Factorization (MF) & $0.3895$ &$0.2300$ &$0.4197$& $0.5585$ & $0.7946$ \\
\hline
\hline
CBF - TF-IDF (Description) & $0.2751$ &$0.1503$ &$0.2777$& $0.3634$ & $0.5494$ \\
CBF - BERT base (Description) & $0.3024$ &$0.1807$ &$0.3050$& $0.3799$ & $0.5668$ \\
CBF - OpenAI (Description) & $0.3113$ &$0.1761$ &$0.3233$& $0.4266$ & $0.6093$ \\
CBF - SBERT (Post) & $0.2865$ &$0.1317$ &$0.3109$& $0.4281$ & $0.6545$ \\
CBF - BERT base (Post) & $0.3140$ &$0.1598$ &$0.3446$& $0.4776$ & $0.6651$ \\
CBF - BERT large (Post) & $0.3168$ &$0.1637$ &$0.3436$& $0.4795$ & $0.6674$ \\
CBF - BERTweet base (Post) & $0.3154$ &$0.1570$ &$0.3516$& $0.4918$ & $0.6700$ \\
CBF - OpenAI (Post) & $0.3195$ &$0.1642$ &$0.3484$& $0.4815$ & $0.6823$ \\
\hline
\hline
MF + CBF OpenAI (Description) & $0.4039$ &$0.2405$ &$0.4491$& $0.5790$ & $0.8093$ \\
MF + CBF BERT base (Post) & $0.4114$ &$0.2449$ &$0.4613$& $0.5966$ & $0.8023 $\\
MF + CBF BERTweet base (Post)  & $0.4221$ &$0.2570$ &$0.4809$& $0.6022$ & $0.8056$ \\
MF + CBF BERT large (Post) & $0.4237$ &$\textbf{0.2593}$ &$0.4832$& $0.6000$ & $0.8059$ \\
\textbf{MF + CBF OpenAI (Post)} & $\textbf{0.4244}$ &$0.2571$ &$\textbf{0.4841}$& $\textbf{0.6063}$ & $\textbf{0.8154}$ \\
\hline
\end{tabular}
\end{center}
\caption{\label{result} Model Performance with different content similarity matrices generated by embedding methods evaluated on $MRR$ and $Recall@K$}
\end{table*}

For CBF, operating a feature-extraction method on subreddit posts results in higher performance than operating the same method on description. For example, the MRR for CBF - BERT base is 0.3140 when using posts and 0.3024 when using description. It can also be observed that given the same information (either posts or information), deep-learning-based feature extraction methods like OpenAI and BERT bring about better performance for CBF than TF-IDF.

\begin{figure}[t]
  \centering
  \includegraphics[width=75mm]{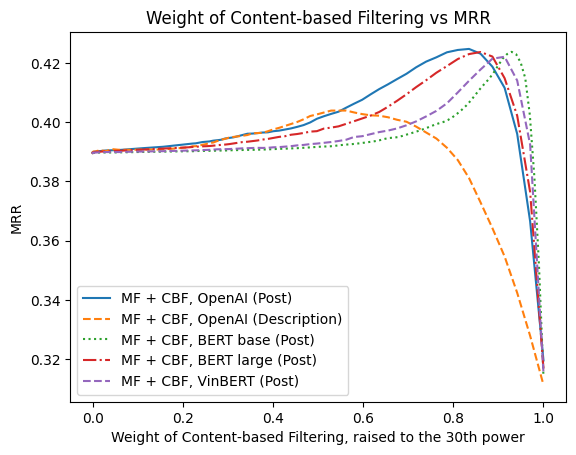}
  \caption{Hybrid Model Performance (MRR) across different values of $\beta$}
  \label{fig:mf_cbf_curve}
  \vspace{-0.1in}
\end{figure}

As our recommendation model combines both MF and CBF, we investigate the effect of hyperparameter $\beta$, which dictates how much CBF contributes to the final prediction. Figure~\ref{fig:mf_cbf_curve} illustrates the performance of the hybrid models on varying $\beta$. When $\beta =  0$, the hybrid model's performance is the same as that of MF. When  $\beta =  1$, the hybrid model's performance is the same as that of CBF. It can be seen from the peak of these curves that this way of linearly combining MF and CBF brings about significant improvement in MRR.

\subsection{Case Studies}
We perform a series of case studies to understand why certain information and methods are more helpful than others in recommending subreddits to users. We present our findings by comparing the behavior of the following models: (1) CBF models using TF-IDF and OpenAI Embedding on Subreddit Descriptions, (2) CBF models using OpenAI Embeddings on Subreddit Descriptions and Posts, and (3) MF model and Hybrid model.

\subsubsection{CBF models using TF-IDF and OpenAI Embedding on Subreddit Descriptions}
\begin{figure}[t]
  \centering
  \includegraphics[width=\columnwidth]{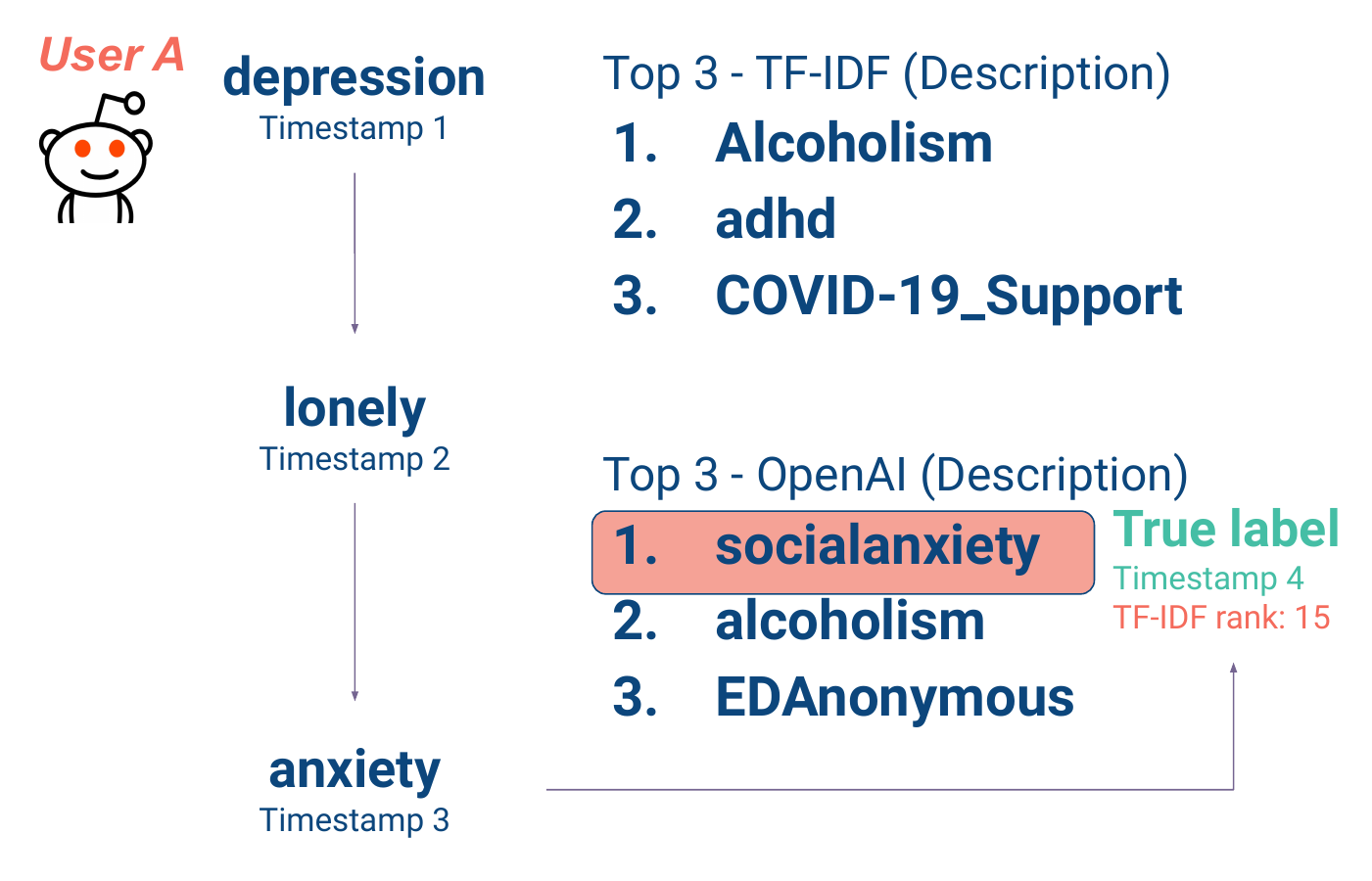}
  \caption{Case Study 1: Top 3 TFIDF Predictions vs. OpenAI Predictions. 
The higher the timestamp, the more recent the interactions between the user under study and the subreddits they engaged with.}
  \label{fig:case_study_1}
\end{figure}

The objective of the first case study is to investigate the impact of different types of embedding methods on the performance of recommendations. To achieve this, we employ TF-IDF and OpenAI Embedding approaches to analyze subreddit descriptions and compare their predictions using content-based filtering (CBF) approaches, as illustrated in Figure~\ref{fig:case_study_1}. Specifically, we consider User A's historically interacted subreddits, which relate to \textit{depression}, \textit{loneliness}, and \textit{anxiety}, respectively, with the ground truth of \textit{socialanxiety}. For CBF models, the content similarity $C$ between historically interacted and ground truth subreddits is crucial for accurate predictions. Hence, we evaluate the similarity scores between them. According to the result, the OpenAI Embedding technique outperforms TF-IDF in learning the representation of subreddits.
Based on the analysis of content similarity matrices of the two approaches, we observe that TF-IDF has low similarity scores among subreddits due to its bag-of-words (BOW) approach, which fails to capture semantic relationships in short texts \cite{naseem2021comprehensive}, such as subreddit descriptions. In contrast, OpenAI Embeddings, which can capture semantic meanings, performs better for encoding the meanings of subreddit descriptions for recommendation tasks.

\subsubsection{CBF models using OpenAI Embeddings on Subreddit Descriptions and Posts}
\begin{figure*}[t]
    \centering
    \includegraphics[width=1.\textwidth]{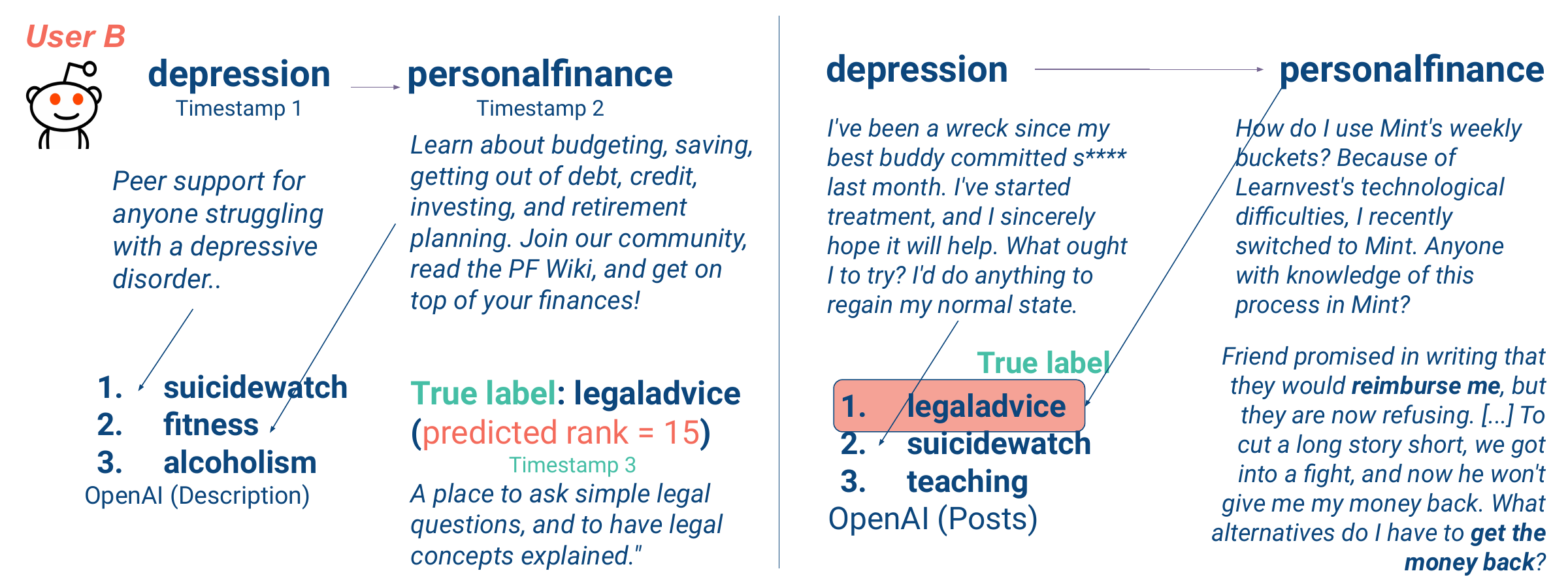}
    \caption{Case Study 2: Top 3 OpenAI Description Predictions (Left) vs. OpenAI Post Predictions (Right). The higher the timestamp, the more recent the interactions between the user under study and the subreddits they engaged with. \textit{Post content has been paraphrased to protect user's privacy}.}
    \label{fig:case_study_2}
\end{figure*}

The second case study aims to investigate the impact of different types of information on the performance and recommendations of CBF models. To achieve this goal, we evaluate OpenAI Embeddings approaches on two types of information, subreddit descriptions, and posts. Figure~\ref{fig:case_study_2} illustrates the predictions using CBF approaches utilizing OpenAI Embeddings on posts and descriptions. Specifically, we examine User B's historical posts, which are in \textit{depression} and \textit{personalfinance}, and the ground truth label is \textit{legaladvice}. To understand the behavior of CBF on these two types of information, we analyze the similarities between historical subreddit interactions of User B and how the ground truth label is correlated with these subreddits. Our analysis shows that using OpenAI Embeddings on subreddit posts can capture strong relationships between \textit{personalfinance} and \textit{legaladvice}, where many \textit{legaladvice} posts are related to financial information. However, when only using subreddit descriptions of \textit{legaladvice}, which is  "A place to ask simple legal questions, and to have legal concepts explained.", the model fails to capture this relationship.
Furthermore, as shown in Table ~\ref{result}, the use of subreddit posts as representations for communities generally exhibits higher performance across most metrics when compared to using community descriptions. The reason is that subreddit descriptions contain less information than posts describing only the general purpose of the subreddit. In contrast, using subreddit posts can accurately learn the representations of the subreddits. Therefore, among the two types of information, using subreddit posts to represent subreddits helps models achieve better performance.

\subsubsection{MF vs MF + CBF model using OpenAI Embeddings on Subreddit Discourses}
\begin{figure}[t]
    \centering
    \includegraphics[width=\columnwidth]{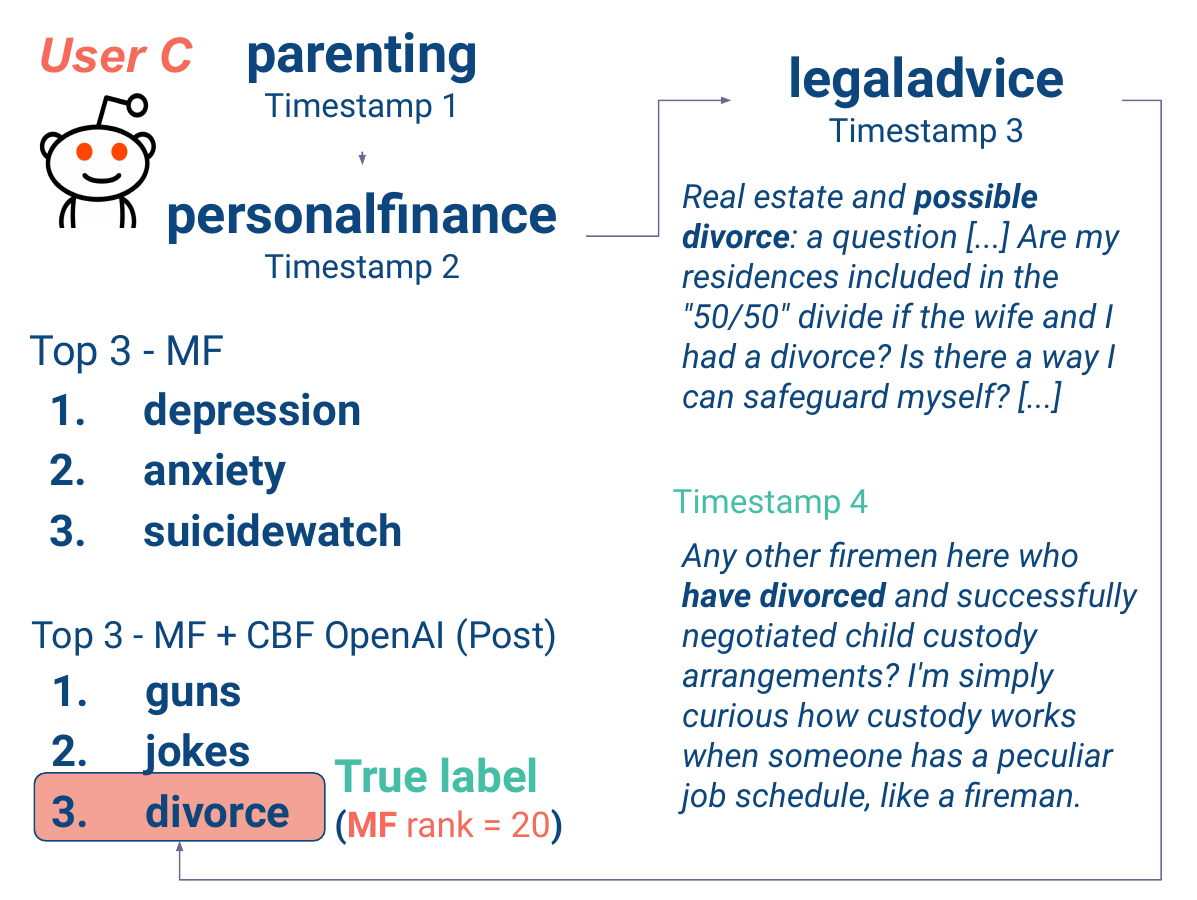}
    \caption{Case Study 3: Top 3 MF Predictions vs. Top 3 MF + CBF Post Predictions. The higher the timestamp, the more recent the interactions between the user under study and the subreddits they engaged with. \textit{Post content has been paraphrased to protect user's privacy}.}
    \label{fig:case_study_3}
\end{figure}

The objective of the third study is to investigate the performance improvement achieved by combining MF and CBF. Specifically, we aim to explore how the use of discourse embeddings to generate content similarity matrices among subreddits can address challenges encountered by the MF approach. To this end, we evaluate the MF and MF + CBF approaches using OpenAI Embeddings on posts. The predictions generated by the two models are presented in Figure \ref{fig:case_study_3}.

\begin{figure}[t]
    \centering
    \includegraphics[width=\columnwidth]{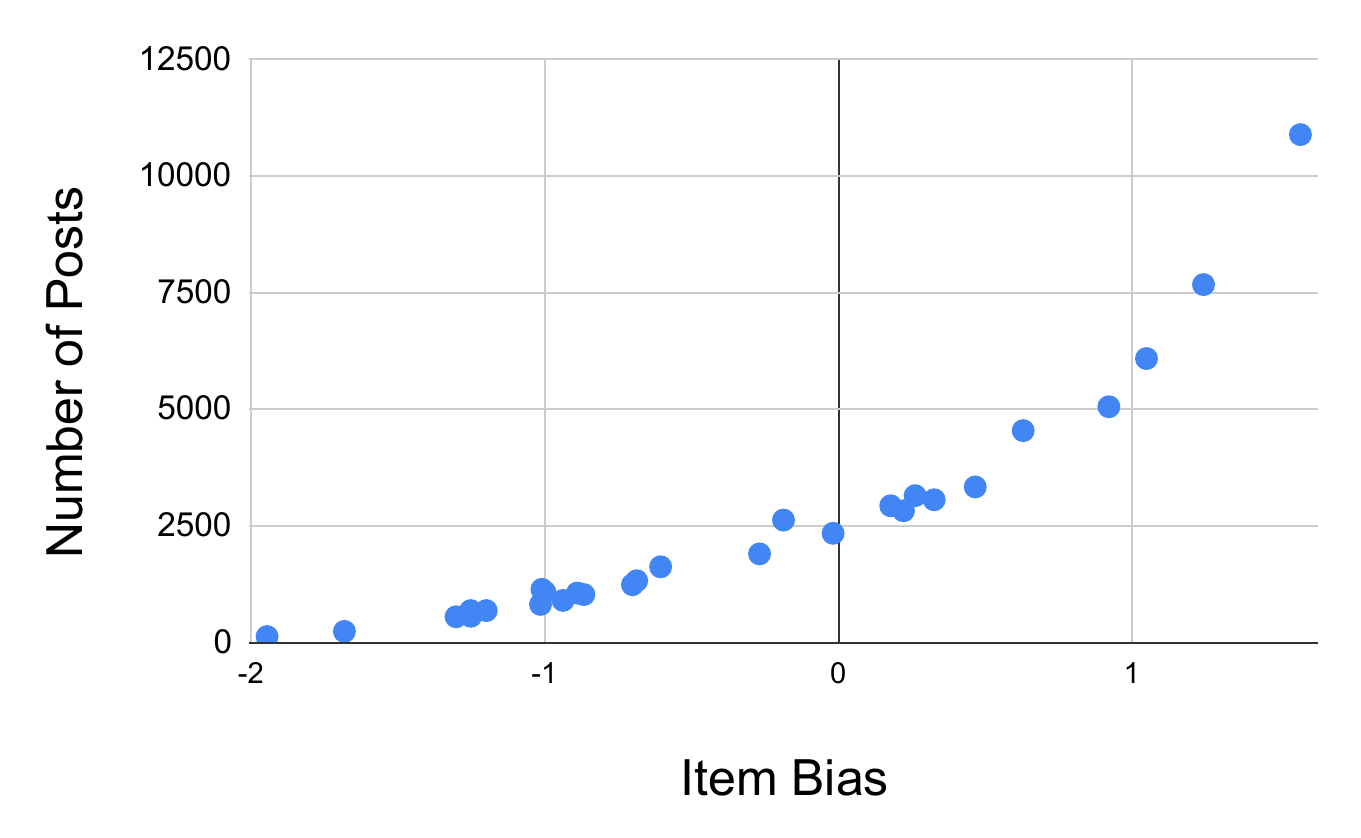}
    \caption{Item Biases values learned from MF vs. Subreddits' number of posts}
    \label{fig:itembiasas}
\end{figure}

We further examine the construction of scores using MF for this case study. The scores values are generated using latent features $P$, $Q$, $\mu$, $b^{(p)}$, and $b^{(q)}$, representing user, item features, global average, user, and item biases, respectively. However, due to the imbalance in the dataset, there are more posts in some subreddits than others, leading to a cold start problem for the MF approach to accurately learn communities with a small number of examples. In this case study, MF fails to generate correct predictions for the \textit{divorce} community due to the limited number of posts available. Additionally, MF is biased towards subreddits with more posts, as reflected by the $b^{(q)}$ values that have strong correlations with the number of posts in the subreddit communities, as depicted in Figure \ref{fig:itembiasas}.

We demonstrate that the top three predictions generated by MF are the subreddits with the highest item biases compared to other subreddits, which are also the ones with the most posts. However, as \textit{divorce} only accounts for $0.78\%$ of the dataset, the performance of MF is limited. By utilizing OpenAI Embeddings on Subreddit Discourses to represent subreddit communities, we can integrate semantic information into the prediction process, thereby overcoming the cold start problem encountered by MF. Furthermore, this approach captures the relationships between the target recommended subreddit, historically interacted communities and semantic similarities. In this case, the most similar subreddits to \textit{personalfinance} are \textit{legaladvice} and \textit{divorce}, while the most similar subreddits to \textit{parenting} are \textit{autism} and \textit{divorce}.


Overall, we showcase that integrating semantic information into MF can address the cold start problem, and combining MF with CBF using discourse embeddings can make better recommendations.

\section{Conclusion}
\label{sec:conclusion}

This study aimed to investigate the effectiveness of different types of discourse embeddings when integrated into content-based filtering for recommending support groups, particularly in the mental health domain. Our findings showed that the hybrid model, which combined content-based filtering and collaborative filtering, yielded the best results. Moreover, we conducted an extensive case study to demonstrate the interpretability of our approach's predictions. 

Previous studies have brought to light the use of past behaviors to make more accurate recommendations in mental health \cite{valentine2022recommender}. They also emphasize effective communication between the recommender system and the user as an essential factor for users' proper understanding of mental health in general as well as in their own journey \cite{valentine2022recommender}. Through promising prediction accuracy and interpretability, we believe that this method can serve as a valuable tool to support individuals, particularly those with mental health concerns, to share and seek help regarding their issues.

\section*{Limitations}
\label{sec:limitations}
In our current project, we have not taken into account the temporal information that treats the historical behavior of users as a sequence of actions. Thus, the model may not capture how user behaviors change over time. To ensure full support to users in need, we recommend that future work should address this limitation by considering users' historical behaviors as a sequence of actions. Moreover, although our pre-trained models achieved significant results without fine-tuning discourse embeddings, we suggest that fine-tuning these models can enhance performance by capturing the nuances of the datasets' distribution and contexts. Furthermore, conducting a detailed comparison of additional open-source Large Language Models (LLMs) would provide more comprehensive insights into their performance.
Additionally, in addition to analyzing the efficiency of different models, it is crucial to evaluate the cost associated with implementing these models. Therefore, future work should consider both fine-tuning and evaluating additional LLMs, while also taking into account the costs of utilizing these models.

\section*{Acknowledgement}
This work was supported by NSF IIS-2119531, IIS-2137396, IIS-2142827, CCF-1901059, and ONR N00014-22-1-2507.

\bibliography{bibliography}
\bibliographystyle{style/acl_natbib}

\end{document}